# Diffuse Soft X-ray Emission from Several Nearby Spiral Galaxies

W. Cui[1], W. T. Sanders, and D. McCammon

Department of Physics, University of Wisconsin-Madison, Madison, WI 53706.

S. L. Snowden

USRA/Laboratory for High Energy Astrophysics,
NASA/Goddard Space Flight Center, Greenbelt, MD 20771

AND

D. S. Womble

California Institute of Technology, Astronomy 105-24, Pasadena, CA 91125

ABSTRACT

We observed several nearby face-on spiral galaxies with the *ROSAT* PSPC to study their 0.1-2.0 keV diffuse emission. After the exclusion of resolved discrete sources, there is unresolved X-ray emission in all the galaxies observed. Since this emission is a combination of diffuse emission and a contribution from unresolved point sources, it represents an upper limit to the truly diffuse soft X-ray emission.

The derived upper limits on the diffuse emission can be interpreted in terms of upper limits to the average intensity of a putative hot halo. They can also be used to derive limits to the total energy radiated by hot gas in the observed galaxies as a function of its temperature for various assumed absorbing geometries. Beyond the equivalent solar radius (the radius at which the Sun would be in the observed galaxies), the temperature of hot gas radiating more than 30% of the total supernova power in the galaxies must be less than $10^{6.1}$ K if it is located within the disk with an assumed absorbing overburden of $3 \times 10^{20}$ cm$^{-2}$, or less than $10^{5.9}$ K if it lies in an unabsorbed halo.

## 1 INTRODUCTION

The diffuse soft X-ray background provides evidence for the existence of million-degree gas in our own Galaxy, but the spatial distribution and filling factor in the Galactic disk are still unknown. Some theoretical models favor isolated hot bubbles with a filling factor in the Galactic disk of around 10% with the Sun by chance inside of one (e.g., Slavin 1989; Slavin & Cox 1992, 1993). Others suggest hot gas filling factors larger than 90% (e.g., McKee & Ostriker 1977), with the Sun in the "typical" environment. The interaction of supernovae with the interstellar medium and most probably the interstellar dynamics regulating star formation are profoundly different in these two types of models. Unfortunately, million-degree gas is primarily observable in soft X-rays, which makes it very difficult (if not impossible) to directly study the global properties of these hot phases in our own galaxy because of heavy interstellar absorption of the soft X-rays in the Galactic disk. The best route to an observational answer to these questions therefore appears to be the study of diffuse emission from other normal spiral galaxies that are located at high galactic latitudes. Here the interstellar medium in our Galaxy is partially transparent, and soft X-ray emission from within and above the disks of these galaxies can be detected.

---

[1] Present address: Center for Space Research, Room 37-571, MIT, Cambridge, MA 02139

Very good sensitivity to large scale-height halo emission can be obtained by looking at edge-on galaxies (e.g., Bregman & Piddis 1994; Pietsch et al. 1994; Wang et al. 1995), but such observations cannot detect emission within the disks, and their interpretations are subject to the large uncertainties in the absorbing geometry for cooler halos or fountains with small scale heights. Observations of face-on spirals, on the other hand, offer high sensitivity to low-energy emission within the disks, and guarantee that any halo or fountain emission from the near side of the galaxy is absorbed only by the foreground column from our Galaxy.

A study of this type was first made in the analysis of *Einstein* IPC observations of M101 (McCammon & Sanders 1984), where upper limits were derived for the diffuse X-ray emission. Such limits are useful in constraining supernova rates and interstellar densities (Cox & McCammon 1986).

We observed several nearby face-on spiral galaxies with the *ROSAT* PSPC. Compared with the *Einstein* IPC, the *ROSAT* PSPC had larger effective area, a larger field of view, lower internal background, and better spatial and spectral resolution, making it an excellent instrument for studying extended X-ray emission. In practice, it is impossible to detect only diffuse emission since there can always be some contribution due to unresolved discrete sources associated with the galaxies. The results therefore must be interpreted conservatively as upper limits on the diffuse emission, which then lead to upper limits on the amount of X-ray emitting hot gas and on the amount of total energy radiated by it.

A fundamental difficulty with this type of observation is that the disk of the observed galaxy is partially opaque at low X-ray energies, and therefore shadows any diffuse background flux originated beyond it. Such a shadow could partially or completely compensate for the increase in surface brightness produced by emission from the galaxy. Unfortunately, the intensity of the extragalactic diffuse background currently is poorly constrained, so this correction produces large uncertainties in the derived emission, despite the high quality and sensitivity of the *ROSAT* observations. Extragalactic shadowing measurements have, in fact, been the primary tool for attempts to measure the diffuse background intensity (Margon et al. 1974; McCammon 1971; McCammon et al. 1976; Barber & Warwick 1994), and in a companion paper (Cui et al. 1995, Paper 2 hereafter) we use these same data to derive limits on the 1/4 keV extragalactic flux. Of course, such measurements suffer from the converse problem of an unknown amount of emission in the target galaxy partially filling in the shadow, but radial variation in both emission and absorption can be combined with existing limits on the extragalactic flux to simultaneously obtain useful results for both. As shown in Section 4 below, absorption by foreground Galactic gas and possible absorption by an overburden of cool gas in the target galaxy conspire to weaken the signal from $10^6$ K disk gas like that surrounding the Sun to the point where uncertainties in the shadowing correction preclude placing useful limits on its filling factor. This important question can still be answered with these data, once the diffuse background intensity has been better determined by other measurements. In Sections 2 and 3, we present the X-ray and H I data, followed in Section 4 by a description of the data reduction and analysis methods. The results are discussed in Section 5, and we conclude in Section 6 by summarizing the main results of this work.

## 2 X-RAY DATA

The strong Galactic interstellar absorption of soft X-rays limited us to looking in directions with low foreground H I column density in order to achieve adequate sensitivity to the soft X-ray emission of the target galaxies. Face-on galaxies were chosen to maximize sensitivity to X-ray emission originating within the disks. The choice of face-on galaxies also facilitates efficient point source removal. Some basic data on the observed galaxies and their lines of sight are presented in Table 1. For comparison we reanalyzed archival *ROSAT* PSPC observations of M101 in the same way as other galaxies.

Table 1: Basic Data

| Name | Type[a] | Galactic Longitude[b] | Galactic Latitude[b] | Distance[a] (Mpc) | Maj. Dia. / Min. Dia.[c] | Galactic $N_{H\,I}$[d] ($10^{20}$ cm$^{-2}$) |
|---|---|---|---|---|---|---|
| NGC 3184 | Sc  | 178 | 56 | 12.1 | 8.5′ / 7.8′ | 1.18 |
| M101     | Sc  | 101 | 60 | 7.4  | 28′ / 28′   | 1.13 |
| NGC 4395 | Sd  | 162 | 82 | 6.1  | 16′ / 13′   | 1.38 |
| NGC 4736 | Sab | 123 | 76 | 6.9  | 14′ / 12′   | 1.44 |
| NGC 5055 | Sbc | 106 | 74 | 11.0 | 15′ / 9′    | 1.27 |

[a] Sandage and Tammann 1987. Distances were derived assuming $H_0$ = 50 km$^{-1}$ s$^{-1}$ Mpc$^{-1}$.
[b] In degrees.
[c] Nilson 1973.
[d] Dickey and Lockman 1990.

We adopted the data reduction procedure and the definition of standard energy bands suggested by Snowden et al. (1994a) for analyzing extended X-ray emission. Table 2 shows the total exposure time and the "good" time remaining after excluding time intervals with the worst contamination by scattered solar X-rays and short-term enhancements. For convenience, both the R12 (PI channels 8-41) and R12L (PI channels 11-41) bands are referred to as the "1/4 keV band", the R45 band (PI channels 52-90) as the "3/4 keV band" and the R67 band (PI channels 91-201) as the "1.5 keV band". Because the observations of M101 and NGC 4736 were carried out in the high gain mode, we analyzed them in the R12 band, while others in the R12L band, to study their 1/4 keV emission. For a reasonable source spectrum, the surface brightness ratio of R12 to R12L is about 1.1, depending only slightly on the spectrum. NGC 3184 was observed partially in the high gain mode and partially in the low gain mode, with the two observations separated by about six months. We chose to process the two data sets separately in the R12L band, using the detector maps appropriate for each gain setting (see Cui 1994 for more detail), and then combined the results

Table 2: Exposure Time and "Good" Time

| Name | Exposure (ksec) | "Good" Time (ksec) |
| --- | --- | --- |
| NGC 3184 | 18.9 | 13.5 |
| M101 | 35.7 | 20.5 |
| NGC 4395 | 11.1 | 8.7 |
| NGC 4736 | 95.7 | 23.7 |
| NGC 5055 | 12.5 | 11.0 |

## 3  H I DATA

For studies of diffuse soft X-ray emission, it is essential to be able to estimate absorbing column densities in and toward the target galaxies. The most practical way is to estimate neutral hydrogen (H I) column densities from the intensity of the 21 cm H I emission line, and then to calculate the total X-ray absorption of material associated with the H I. The galaxies under study are small (major diameters less than 28'), so interferometric resolution H I data is essential for spatial correlation studies between the X-ray and H I images. Three of our target galaxies were observed in the 21 cm H I emission line with either the Very Large Array (VLA) or the Westerbork Synthesis Radio Telescope (WSRT).

NGC 3184 was observed by Womble (1993) using the VLA with angular resolution of ~1', comparable to that of the *ROSAT* PSPC at 1/4 keV. Here we simply rebinned and precessed the obtained H I column density image to the epoch of J2000 to match the *ROSAT* X-ray images. The typical rms noise level is approximately $2 \times 10^{19}$ cm$^{-2}$. This face-on galaxy shows a normal distribution of H I, including a sharp edge at 2.2 optical-galaxy radii (Womble 1993).

NGC 4736 was also observed with the VLA, and an H I column density image with angular resolution of ~1' was made available to us by Braun (1994). It has an rms noise level less than $2 \times 10^{19}$ cm$^{-2}$. Again we rebinned and precessed the image to match the X-ray images. This galaxy has a very unusual H I distribution (Bosma, van der Hulst, & Sullivan 1977). There are at least four distinct regions. A bright central core extends out to a radius of about 15"; between radii of 15" and 50" a small spiral structure can be clearly seen; the latter is surrounded by a dense ring structure between radii of 1' and 3'; a fainter outer ring shows up at the most outside edge. The two rings touch each other on the east and west sides and appear to be separated by a gap on the north and south sides.

M101 was observed with the WSRT with angular resolution of ~30" and an rms noise less than $1.8 \times 10^{19}$ cm$^{-2}$ (van der Hulst & Sancisi 1988; Kamphuis, Sancisi, & van der Hulst 1991).

We do not have interferometric H I data for NGC 4395 and NGC 5055.

## 4. ANALYSIS

To treat all the galaxies uniformly, we binned the data in concentric rings such that there are always four rings within the largest visually determined extent of each galaxy, as listed in the UGC catalog (see Table 1). This turns out to be roughly 1.25 times the average isophotal diameter at the level of 25$^{th}$ B-magnitude per square arcsecond. The angular width of each ring is the same for a given galaxy, but differs from galaxy to galaxy. We defined the average X-ray surface brightness between the 8$^{th}$ and 16$^{th}$ rings as a baseline level (BL) representing the sum of foreground and background diffuse X-ray emission.

### 4.1. *The Cleaned X-Ray Images*

The cleaned *ROSAT* count maps in the 1/4 keV band were obtained by distributing the observed events in R12 or R12L bands into $512 \times 512$ pixel images with a pixel size of 15" in sky coordinates, and subtracting from them the corresponding count maps of non-cosmic X-ray background (Snowden et al. 1994a). Appropriate detector maps, along with attitude information and live times, were used to create efficiency-weighted exposure maps that were then used to make exposure-corrected X-ray count rate images. As an example, Fig. 1 shows the 1/4 keV image of NGC 4736. Note that the image was smoothed with a 2-dimensional Gaussian filter with FWHM of 45". In a similar manner we produced the 3/4 keV and 1.5 keV images for each of the observed galaxies. Those of NGC 4736 are shown in Figures 2 and 3, respectively.

### 4.2. *Point Source Detection*

Since we are interested only in the diffuse X-ray emission, we would like to eliminate contributions from the foreground Galactic X-ray sources and from discrete X-ray sources associated with the galaxies.

This was not practical, but we removed point sources to the faintest practical flux limit to minimize the excess fluctuations caused by them.

We used a simple sliding-box technique with a fixed detection significance limit of 3.5 σ, which guarantees less than one spurious source within the inner part of the field of view (r < 20'). We carried out point source detection separately for the 1/4 keV band and for the 0.5-2 keV band. For each source detected, a mask was created to blank out the region around the source with a radius of 1.5 times the 90% power radius of the point spread function (PSF). The small circles shown in Figures 1-3 indicate the positions of the sources detected and the regions of the images removed from further analysis.

Although the 1/4 keV band is less sensitive than the harder bands for detecting extragalactic X-ray sources, due to greater foreground absorption, it is reasonable to use it for point source detection in studying the 1/4 keV diffuse emission. The extragalactic sources directly behind the galaxies can hardly shine through in this soft energy band, and their contribution to the off-galaxy baseline level can be calculated and then subtracted by using the known LogN-LogS relation, providing source detection flux limits are known. The lower point source detection sensitivity in this band does leave more extragalactic sources unresolved in regions surrounding the galaxies, thus increasing the baseline fluctuations, and additional sensitivity could be obtained by also searching in the 3/4 and/or 1.5 keV bands. However, the gain would be small since large off-galaxy areas can be averaged to minimize statistical uncertainties of the baseline level.

To study harder diffuse X-ray emission, we performed the point-source detection in the 0.5-2.0 keV band following the same steps as discussed above.

### 4.3. *1/4 keV Unresolved Emission*

Radial profiles of the observed X-ray surface brightness were obtained for the observed galaxies, and are given in Table 3 for the inner eight rings along with the baseline level. To facilitate comparisons among the galaxies, which have different amounts of foreground absorption, the observed surface brightnesses are corrected for foreground absorption as described below.

For each galaxy, the foreground H I column density is determined from Dickey & Lockman (1990), and is listed in Table 1. Absorption by helium and metals associated with the diffuse H II layer must also be included. This layer has an average column density of $7 \times 10^{19}$ csc|b| H II cm$^{-2}$ (Reynolds 1991). If we assume the helium is 50% singly ionized and 50% neutral, the effective 1/4 keV absorption cross section of the H II layer is then about 64% of that for entirely neutral gas (Snowden et al. 1994b). Combining the H II layer absorption with that of the foreground H I, we calculated the average foreground 1/4 keV transmission, assuming an X-ray spectrum of equilibrium $10^6$ K plasma and the absorption cross sections of Morrison & McCammon (1983, MM hereafter). The results are shown in Table 4. However, the absorbing column densities of the H II layer are still quite uncertain in most directions, since there are only limited number of measurements, so we have also included in Table 4 the average foreground soft X-ray transmission of only the neutral gas in our Galaxy. After correction for foreground absorption (with the H II layer taken into account), the X-ray radial profiles are shown in Fig. 4. Note that fluctuations in the off-galaxy surface brightness are caused partly by the brightest undetected sources.

Table 3: Observed 1/4 keV Surface Brightness*

| Ring | NGC 3184 | NGC 4736 | M101 | NGC 4395 | NGC 5055 |
|---|---|---|---|---|---|
| 1 | 1555 ± 285 | 2326 ± 288 | 2422 ± 78 | 973 ± 137 | 1644 ± 236 |
| 2 | 1407 ± 122 | 1358 ± 58 | 1804 ± 33 | 1032 ± 70 | 1387 ± 72 |
| 3 | 1106 ± 78 | 1030 ± 39 | 1612 ± 24 | 913 ± 53 | 1060 ± 48 |
| 4 | 900 ± 60 | 912 ± 34 | 1582 ± 20 | 853 ± 44 | 1002 ± 41 |
| 5 | 1051 ± 57 | 996 ± 31 | 1594 ± 19 | 935 ± 41 | 973 ± 35 |
| 6 | 1115 ± 55 | 996 ± 28 | 1619 ± 20 | 932 ± 36 | 1014 ± 32 |
| 7 | 1079 ± 52 | 956 ± 26 | 1616 ± 20 | 981 ± 36 | 929 ± 28 |
| 8 | 987 ± 47 | 946 ± 24 | 1610 ± 17 | 919 ± 34 | 922 ± 27 |
| BL | 1060 ± 12 | 945 ± 7 | 1630 ± 5 | 935 ± 10 | 965 ± 8 |

*All intensities are in units of $10^{-6}$ count s$^{-1}$ arcmin$^{-2}$.

Table 4: Foreground 1/4 keV Transmission

| Case | NGC 3184 | NGC 4736 | M101 | NGC 4395 | NGC 5055 |
|---|---|---|---|---|---|
| With H II | 0.224 | 0.180 | 0.215 | 0.197 | 0.210 |
| No H II | 0.322 | 0.252 | 0.322 | 0.275 | 0.300 |

We cannot simply subtract the baseline level from the on-galaxy surface brightness to get emission radial profiles because the baseline level also includes a contribution from the extragalactic X-ray background, which is almost completely absorbed toward the galaxy. This soft X-ray shadowing by the galaxy must be corrected. Moreover, the increasing size of PSF with distance from the center of the detector, along with the reduced effective area due to vignetting, results in more point sources detected and removed in the more sensitive central part of the field of view than in the outer region for a given detection significance level. The resulting non-uniform source removal can potentially introduce artificial radial structures which must also be corrected. In the following two sub-sections, we discuss in detail how we made these corrections, since it is a very important part of the analysis procedure.

4.3.1. *Correction For Non-Uniform Source Removal and Shadowing of the Known Extragalactic Background*

The effects of non-uniform source removal can be calculated by using the LogN-LogS relation derived from a deep ROSAT survey by Hasinger et al. (1993, H93 hereafter). They also found that the sources with unabsorbed 0.5-2.0 keV fluxes between 0.25 and $4 \times 10^{-14}$ ergs cm$^{-2}$ s$^{-1}$ (*ROSAT* sources, hereafter) contributed about $130 \times 10^{-6}$ count s$^{-1}$ arcmin$^{-2}$ in the R12 band, and had a power law spectrum of 7.8 E$^{-0.96}$ keV cm$^{-2}$ s$^{-1}$ sr$^{-1}$ keV$^{-1}$ with an absorbing column density of $8.7 \times 10^{19}$ cm$^{-2}$. After correction for the Galactic foreground absorption (using the absorption cross sections of MM), the *ROSAT* sources produce about 265 and $294 \times 10^{-6}$ count s$^{-1}$ arcmin$^{-2}$ in the R12L and R12 band, respectively.

Since the absolute normalization of the LogN-LogS relation is still uncertain, we decide to use it only in a relative sense. For each ring, the integrated surface brightness due to sources having fluxes below $S^i_{lim}$ but above the H93 limit, the remaining *ROSAT* sources, is given by

$$I^i_{egk} = I_{ROSAT} \, T^i \times \frac{\int_{0.25}^{S^i_{lim}} S \times N(S) \, dS}{\int_{0.25}^{4.0} S \times N(S) \, dS}, \quad (1)$$

where N(S) is the number of sources per square degree per unit flux at flux S (LogN-LogS relation); S is the 0.5 - 2.0 keV source flux in the units of $10^{-14}$ ergs cm$^{-2}$ s$^{-1}$; I$_{ROSAT}$ is the integrated X-ray surface brightness of the ROSAT sources (see above); $S^i_{lim}$ is the *unabsorbed* source-detection flux limit at the i$^{th}$ ring; and T$^i$ is the X-ray transmission at the i$^{th}$ ring. For the i$^{th}$ ring, the *unabsorbed* source-detection flux limit ($S^i_{lim}$) is calculated by dividing the source-detection flux limit ($S^i_{det}$, already corrected for foreground Galactic absorption) by the X-ray transmission (T$^i$), i.e., $S^i_{lim} = S^i_{det} / T^i$. The average 1/4 keV transmissions of the observed galaxies were calculated ring by ring by using the H I maps when available (assuming a source spectrum $\propto$ E$^{-0.96}$ and the absorption cross sections of MM). For galaxies without H I data, we assume an average transmission of 5% out to the optical radius (the outer edge of ring 4) and complete transparency beyond it. The radial profiles of the average 1/4 keV X-ray transmission in NGC 3184, M101 and NGC 4736, are shown in Fig. 5, along with the assumed transmission curves for NGC 4395 and NGC 5055.

Using equation (1), we calculated the integrated surface brightness of the remaining *ROSAT* sources for the inner 8 rings of each galaxy, as well as an area-weighted average surface brightness for each observation between the 8$^{th}$ and 16$^{th}$ rings, representing the contribution to the baseline level from unremoved point sources. The results are shown in Table 5a and Table 5b for cases with and without correction for foreground absorption by gas associated with the H II layer. Note that the different assumptions for foreground absorption affect the value for the unabsorbed flux limit used in equation (1). Assuming no source clustering, the rms uncertainty in $I^{\,i}_{egk}$ due to statistical fluctuations in the number of sources is calculated as

$$\sigma^{\,i}_{egk} = I_{ROSAT}\, T^{\,i} \times \frac{\sqrt{\int_{0.25}^{S^{\,i}_{lim}} S^{\,2} \times N(S)\, dS}}{\sqrt{A^{\,i}} \times \int_{0.25}^{4.0} S \times N(S)\, dS}, \quad (2)$$

where $A^i$ is the solid angle subtended by the $i^{th}$ ring in units of degree squared. These values are also shown in Table 5a and Table 5b.

Table 5a: 1/4 keV Contribution from the Remaining *ROSAT* Sources[*]

| Ring | NGC 3184 | NGC 4736 | M101 | NGC 4395 | NGC 5055 |
|------|----------|----------|------|----------|----------|
| 1  | 17 ± 57  | 20 ± 39  | 63 ± 50  | 18 ± 32  | 18 ± 35  |
| 2  | 12 ± 25  | 80 ± 70  | 61 ± 28  | 18 ± 20  | 18 ± 20  |
| 3  | 14 ± 19  | 120 ± 68 | 77 ± 24  | 18 ± 14  | 18 ± 14  |
| 4  | 31 ± 29  | 150 ± 63 | 125 ± 29 | 18 ± 11  | 18 ± 11  |
| 5  | 85 ± 55  | 208 ± 69 | 144 ± 27 | 277 ± 82 | 263 ± 81 |
| 6  | 165 ± 79 | 238 ± 65 | 177 ± 28 | 283 ± 72 | 270 ± 71 |
| 7  | 211 ± 84 | 270 ± 63 | 228 ± 31 | 289 ± 65 | 277 ± 64 |
| 8  | 225 ± 78 | 279 ± 58 | 320 ± 38 | 294 ± 59 | 282 ± 58 |
| BL | 268 ± 73 | 320 ± 33 | 366 ± 11 | 321 ± 28 | 310 ± 30 |

[*]All intensities are in units of $10^{-6}$ count s$^{-1}$ arcmin$^{-2}$.

Subtracting this contribution (for the case with the H II layer taken into account, as listed in Table 5a) ring by ring from the foreground-absorption-corrected surface brightness (as shown in Fig. 4), the corrected radial emission profiles are shown in Fig. 6. Compared with Fig. 4, the depth of absorption features seen at the outer edge of the disks are much reduced, mostly due to the removal of the *calculated* effects of the shadowing of the *known* extragalactic sources that were resolved in the deep ROSAT survey.

Of course, any artificial radial structures produced by non-uniform source removal should also have been removed, but these effects are hardly noticeable in this energy band, and will be more apparent in the hard energy bands. Another important thing to notice is that the error bars in Fig. 6 are larger than those in Fig. 4 due to the statistical fluctuation in the number of remaining sources. For this reason we prefer to detect and remove as many of the extragalactic sources as possible at the very beginning in order to minimize fluctuation in X-ray surface brightness introduced by them.

Table 5b: 1/4 keV Contribution from the Remaining *ROSAT* Sources[*]

| Ring | NGC 3184 | NGC 4736 | M101 | NGC 4395 | NGC 5055 |
|------|----------|----------|------|----------|----------|
| 1 | 17 ± 52 | 20 ± 37 | 60 ± 45 | 18 ± 32 | 18 ± 33 |
| 2 | 12 ± 24 | 77 ± 64 | 59 ± 25 | 18 ± 19 | 18 ± 19 |
| 3 | 13 ± 18 | 113 ± 62 | 74 ± 22 | 18 ± 13 | 18 ± 13 |
| 4 | 31 ± 27 | 139 ± 56 | 118 ± 26 | 18 ± 10 | 28 ± 10 |
| 5 | 81 ± 50 | 187 ± 60 | 137 ± 25 | 253 ± 72 | 234 ± 69 |
| 6 | 149 ± 68 | 209 ± 56 | 167 ± 25 | 261 ± 64 | 243 ± 62 |
| 7 | 184 ± 70 | 235 ± 54 | 215 ± 28 | 268 ± 57 | 251 ± 55 |
| 8 | 192 ± 64 | 246 ± 50 | 297 ± 33 | 274 ± 53 | 258 ± 51 |
| BL | 235 ± 61 | 296 ± 30 | 350 ± 9 | 308 ± 26 | 293 ± 27 |

[*]Same as in Table 5a, but without correction for absorption by gas in the H II layer.

### 4.3.2. *Correction for Shadowing of the Unknown Extragalactic X-Ray Background*

Having subtracted the extragalactic background that we know how to quantify, we now account for the sources below the H93 threshold. The on-galaxy surface brightness (rings 1-4) in Fig. 6 is given by

$$I^i_{cor} = I^i_{gal} + I_{egu} \bullet T^i + I_{MW}, \quad (3)$$

where $I_{egu}$ is the surface brightness of the *unknown* 1/4 keV extragalactic background, $I^i_{gal}$ is the surface brightness of the unresolved 1/4 keV X-ray emission at the i$^{th}$ ring of an observed galaxy, $T^i$ is the soft X-ray transmission at that ring, and $I_{MW}$ is the foreground X-ray surface brightness; and the baseline level is given by

$$I_{cor}^{base} = I_{egu} + I_{MW}, \quad (4)$$

by subtracting equation (4) from equation (3), we therefore have

$$I_{cor}^{i} - I_{cor}^{base} = I_{gal}^{i} - (1 - T^{i})I_{egu}. \quad (5)$$

Solving for $I_{gal}^{i}$, we obtain

$$I_{gal}^{i} = (I_{cor}^{i} - I_{cor}^{base}) + I_{egu} \times (1 - T^{i}). \quad (6)$$

Note again that

$$I_{cor} = \frac{I_{obs}}{T_f} - I_{rem}, \quad (7)$$

where $I_{obs}$ is the observed surface brightness, $T_f$ is the foreground X-ray transmission, and $I_{rem}$ is the integrated surface brightness of the remaining *ROSAT* sources.

From equation (6) it is obvious that in order to derive upper limits to the diffuse emission from the observed galaxies, we should use an upper limit on the intensity of the *unknown* extragalactic flux ($I_{egu}$). We adopted the best 95% confidence upper limit on the intensity of the 1/4 keV extragalactic background derived from observations of the Small Magellanic Cloud ($I_{SMC}$ hereafter; McCammon 1971; McCammon et al. 1976), 45 keV cm$^{-2}$ s$^{-1}$ sr$^{-1}$ keV$^{-1}$ at 1/4 keV when taking into account absorption by gas associated with the H II layer (McCammon & Sanders 1990, McS hereafter). Converted to the PSPC count rate, this upper limit gives 558 and 621 $\times$ 10$^{-6}$ count s$^{-1}$ arcmin$^{-2}$ in the R12L and R12 bands, respectively, assuming a power law source spectrum $\propto E^{-0.96}$.

This upper limit, however, includes *all* the extragalactic point sources up to effectively infinite source flux, and our data have already been corrected for the extragalactic sources down to the H93 limit. We therefore integrate the LogN-LogS relation from the H93 limit to infinite source flux to get the surface brightness of all the resolved point sources: ~377 and 421 $\times$ 10$^{-6}$ count s$^{-1}$ arcmin$^{-2}$ in the R12L and R12 bands, respectively, representing the X-ray surface brightness of the *known* extragalactic background. Subtracting it from $I_{SMC}$, we obtain an upper limit to the surface brightness of *unknown* extragalactic background equal to 181 and 200 $\times$ 10$^{-6}$ count s$^{-1}$ arcmin$^{-2}$ in the R12L and R12 bands, respectively. Substituting these values into equation (6), and taking a 95% confidence upper limit of the result, we derive the upper limit to the diffuse emission in each ring, shown in Table 6a and Table 6b (only for the inner four rings), for both cases with and without correction for foreground absorption by gas associated with the H II layer. We can also obtain a lower limit to the diffuse emission by setting $I_{egu}$ equal to zero in equation (6), and then taking a 95% confidence lower limit of the result. The results are shown in Table 7a and Table 7b. As can be seen from these results, there is little emission outside of the second ring for almost all of the observed galaxies. Since we do not have H I data for NGC 4395 and NGC 5055, and the limits for these galaxies were derived based on the assumed values for the X-ray transmission, and should be taken with reservation.

Table 6a: 95% Confidence Upper Limit on 1/4 keV Emission*

| Galaxy | 1 | 2 | 3 | 4 | 5 | 6 | 7 | 8 |
|---|---|---|---|---|---|---|---|---|
| NGC 3184 | 4681 | 2867 | 1213 | 143 | 718 | 864 | 598 | 124 |
| NGC 4736 | 10739 | 3328 | 926 | 422 | 764 | 675 | 377 | 293 |
| M101 | 4743 | 1535 | 550 | 309 | 328 | 401 | 303 | 111 |
| NGC 4395 | 1785 | 1543 | 805 | 429 | 414 | 351 | 590 | 252 |
| NGC 5055 | 5502 | 3033 | 1290 | 964 | 393 | 554 | 112 | 60 |

*All intensities are in units of $10^{-6}$ count $s^{-1}$ $arcmin^{-2}$.

Table 6b: 95% Confidence Upper Limit on 1/4 keV Emission*

| Galaxy | 1 | 2 | 3 | 4 | 5 | 6 | 7 | 8 |
|---|---|---|---|---|---|---|---|---|
| NGC 3184 | 3551 | 2095 | 944 | 195 | 574 | 649 | 449 | 116 |
| NGC 4736 | 7782 | 2404 | 745 | 324 | 600 | 527 | 303 | 238 |
| M101 | 3307 | 1170 | 506 | 324 | 327 | 360 | 273 | 106 |
| NGC 4395 | 1402 | 1228 | 699 | 430 | 329 | 281 | 447 | 203 |
| NGC 5055 | 3975 | 2246 | 1028 | 798 | 312 | 420 | 106 | 67 |

*Same as in Table 6a, without correction for absorption by gas in the H II layer.

Table 7a: 95% Confidence Lower Limit on 1/4 keV Emission[*]

| Galaxy | 1 | 2 | 3 | 4 |
|---|---|---|---|---|
| NGC 3184 | 419 | 922 | -117 | -934 |
| NGC 4736 | 5417 | 2004 | 51 | -344 |
| M101 | 3399 | 862 | 18 | -144 |
| NGC 4395 | -621 | 219 | -251 | -483 |
| NGC 5055 | 1726 | 1747 | 372 | 146 |

[*]All intensities are in units of $10^{-6}$ count s$^{-1}$ arcmin$^{-2}$.

Table 7b: 95% Confidence Lower Limit on 1/4 keV Emission[*]

| Galaxy | 1 | 2 | 3 | 4 |
|---|---|---|---|---|
| NGC 3184 | 331 | 682 | -40 | -614 |
| NGC 4736 | 3928 | 1470 | 68 | -217 |
| M101 | 2353 | 661 | 92 | -29 |
| NGC 4395 | -374 | 228 | -107 | -274 |
| NGC 5055 | 1279 | 1294 | 328 | 172 |

[*]Same as in Table 7a, without correction for absorption by gas in the H II layer.

### 4.4. *3/4 keV Unresolved Emission*

As for the 1/4 keV emission, the observed 3/4 keV surface brightnesses in each of the eight inner rings and the baseline level were derived, and are listed in Table 8. Following a small correction for the foreground absorption (assuming an equilibrium thermal spectrum of $10^{6.4}$ K plasma and the absorption cross sections of MM; additional absorption by gas associated with the H II layer is negligible in this energy range), we then corrected for the effects of non-uniform source removal and shadowing of the remainder of the known extragalactic background with the knowledge of 3/4 keV surface brightness of the ROSAT sources (~$32 \times 10^{-6}$ count s$^{-1}$ arcmin$^{-2}$; H93) and X-ray transmission of the galaxies. For galaxies with no H I data available, we assumed an optical depth of 1/4 out to the optical radius, and complete transparency beyond it. The 3/4 keV transmission used in this analysis are shown in Fig. 7. Fig. 8 shows the corrected radial emission profiles.

Table 8: Observed 3/4 keV Surface Brightness[*]

| Galaxy | 1 | 2 | 3 | 4 | 5 | 6 | 7 | 8 | BL |
|---|---|---|---|---|---|---|---|---|---|
| NGC 3184 | 256 ± 105 | 257 ± 53 | 178 ± 32 | 107 ± 20 | 73 ± 15 | 143 ± 19 | 126 ± 18 | 128 ± 16 | 102 ± 4 |
| NGC 4736 | 518 ± 87 | 254 ± 27 | 118 ± 15 | 114 ± 13 | 111 ± 12 | 126 ± 11 | 136 ± 11 | 134 ± 10 | 131 ± 3 |
| M101 | 511 ± 34 | 299 ± 16 | 204 ± 11 | 204 ± 10 | 226 ± 9 | 212 ± 10 | 214 ± 10 | 213 ± 8 | 231 ± 2 |
| NGC 4395 | 342 ± 74 | 219 ± 31 | 103 ± 16 | 88 ± 13 | 126 ± 14 | 130 ± 12 | 114 ± 11 | 138 ± 12 | 130 ± 4 |
| NGC 5055 | 240 ± 85 | 237 ± 30 | 150 ± 18 | 112 ± 14 | 112 ± 12 | 108 ± 11 | 84 ± 10 | 104 ± 10 | 112 ± 3 |

[*]All intensities are in units of $10^{-6}$ count s$^{-1}$ arcmin$^{-2}$.

For the intensity of the 3/4 keV extragalactic background, we adopted an upper limit of 22 keV cm$^{-2}$ s$^{-1}$ sr$^{-1}$ keV$^{-1}$(McS), which corresponds to about $112 \times 10^{-6}$ count s$^{-1}$ arcmin$^{-2}$ in the 3/4 keV band assuming a power law source spectrum $\propto E^{-0.96}$. Subtracting from it the contribution from the *known* extragalactic sources (H93), we derive an upper limit to the intensity of the *unknown* 3/4 keV extragalactic X-ray background, $67 \times 10^{-6}$ count s$^{-1}$ arcmin$^{-2}$. As for the 1/4 keV emission, we derived the upper and lower 95% confidence limits using equation (6), as shown in Table 9 and Table 10, respectively. There is little 3/4 keV emission outside of the second ring.

Table 9: 95% Confidence Upper Limit on 3/4 keV Emission[*]

| Galaxy | 1 | 2 | 3 | 4 | 5 | 6 | 7 | 8 |
|---|---|---|---|---|---|---|---|---|
| NGC 3184 | 375 | 289 | 169 | 65 | 13 | 90 | 67 | 66 |
| NGC 4736 | 611 | 207 | 35 | 23 | 15 | 28 | 35 | 30 |
| M101 | 391 | 135 | 25 | 13 | 30 | 12 | 10 | 2 |
| NGC 4395 | 393 | 185 | 33 | 10 | 35 | 34 | 13 | 39 |
| NGC 5055 | 319 | 221 | 107 | 56 | 34 | 28 | -1 | 17 |

[*]All intensities are in units of $10^{-6}$ count s$^{-1}$ arcmin$^{-2}$.

Table 10: 95% Confidence Lower Limit on 3/4 keV Emission[*]

| Galaxy | 1 | 2 | 3 | 4 |
|---|---|---|---|---|
| NGC 3184 | -13 | 80 | 31 | -28 |
| NGC 4736 | 282 | 99 | -28 | -33 |
| M101 | 273 | 65 | -30 | -32 |
| NGC 4395 | 116 | 58 | -42 | -57 |
| NGC 5055 | 6 | 98 | 24 | -11 |

[*]All intensities are in units of $10^{-6}$ count s$^{-1}$ arcmin$^{-2}$.

### 4.5. *1.5 keV Diffuse X-Ray Emission*

In the 1.5 keV band, the absorption of X-rays by gas associated with the observed galaxies is negligible. With X-ray transmission nearly equal to unity, there is no need for absorption corrections. However, the effects of non-uniform source removal and statistical fluctuations due to the remaining *ROSAT* sources are more apparent, and must be corrected.

Table 11: Observed 1.5 keV Surface Brightness[*]

| Galaxy | 1 | 2 | 3 | 4 | 5 | 6 | 7 | 8 | BL |
|---|---|---|---|---|---|---|---|---|---|
| NGC 3184 | 335 ± 122 | 259 ± 54 | 152 ± 30 | 119 ± 22 | 91 ± 17 | 124 ± 19 | 87 ± 15 | 110 ± 15 | 85 ± 4 |
| NGC 4736 | 396 ± 59 | 213 ± 19 | 106 ± 11 | 75 ± 8 | 83 ± 8 | 97 ± 8 | 113 ± 8 | 89 ± 7 | 103 ± 2 |
| M101 | 300 ± 24 | 206 ± 11 | 167 ± 8 | 155 ± 7 | 146 ± 6 | 164 ± 7 | 159 ± 7 | 145 ± 6 | 169 ± 2 |
| NGC 4395 | 371 ± 78 | 198 ± 30 | 96 ± 15 | 86 ± 13 | 100 ± 12 | 79 ± 10 | 104 ± 10 | 104 ± 10 | 106 ± 3 |
| NGC 5055 | 700 ± 143 | 257 ± 31 | 147 ± 17 | 113 ± 14 | 104 ± 11 | 87 ± 10 | 82 ± 9 | 86 ± 8 | 107 ± 3 |

[*]All intensities are in units of $10^{-6}$ count s$^{-1}$ arcmin$^{-2}$.

The observed 1.5 keV surface brightness profile for each galaxy is listed in Table 11 for the inner eight rings and baseline level, and is also plotted in Fig. 9. By examining the radial profiles, the effects of non-uniform point source removal are apparent: more point sources were detected and removed in the central regions than in the outer areas, so the radial emission profiles show a gradual increase in off-galaxy surface brightness with increasing radius. As before, we correct the effects of non-uniform source removal by using the LogN-LogS relation and the observed surface brightness of the ROSAT sources in this energy band (~$52 \times 10^{-6}$ count s$^{-1}$ arcmin$^{-2}$; H93). The corrected emission radial profiles are shown in Fig. 10. Using equation (6) ($T^i$ = 1), we calculated the 1.5 keV surface brightness. The results are shown in Table 12.

Table 12: The Measured 1.5 keV Surface Brightness[*]

| Galaxy | 1 | 2 | 3 | 4 | 5 | 6 | 7 | 8 |
|---|---|---|---|---|---|---|---|---|
| NGC 3184 | 263 ± 132 | 187 ± 61 | 80 ± 37 | 47 ± 28 | 19 ± 23 | 52 ± 24 | 12 ± 21 | 35 ± 21 |
| NGC 4736 | 317 ± 62 | 134 ± 22 | 27 ± 14 | -7 ± 11 | 1 ± 11 | 12 ± 10 | 24 ± 11 | -2 ± 9 |
| M 101 | 168 ± 28 | 71 ± 14 | 29 ± 11 | 11 ± 10 | -5 ± 9 | 10 ± 9 | 1 ± 9 | -16 ± 8 |
| NGC 4395 | 292 ± 83 | 119 ± 34 | 17 ± 20 | 4 ± 17 | 15 ± 16 | -11 ± 14 | 12 ± 13 | 9 ± 13 |
| NGC 5055 | 619 ± 146 | 176 ± 35 | 66 ± 21 | 29 ± 18 | 17 ± 15 | -2 ± 13 | -9 ± 12 | -9 ± 12 |

[*]All intensities are in units of $10^{-6}$ count s$^{-1}$ arcmin$^{-2}$.

## 5. DISCUSSION

Unresolved emission is apparent in all energy bands for all the galaxies observed, except for NGC 4395 in the 1/4 keV band. The emission is the brightest at the center of each galaxy and fades rapidly toward the outer edges. Snowden & Pietsch (1995) argued that the observed 0.1-1 keV extended central emission in M101 (similar to what we see within the two inner rings in these galaxies) cannot be accounted for by any known classes of X-ray sources, based on the temperatures for 1/4 and 3/4 keV emission derived by using the R2/R1 and R5/R4 ratios and X-ray luminosities in these energy bands. However, the observed extended emission does include an unknown amount of contribution from unresolved point sources in the galaxies, so only represents an upper limit to the truly diffuse emission.

As mentioned in Section 1, one fundamental difficulty with this type of studies is the uncertainty in "shadowing" correction due to our lack of understanding of the extragalactic X-ray background in this energy band. In this paper we derived the upper limits to the 1/4 keV emission from target galaxies by using the SMC upper limit, which was derived from a sounding rocket experiment carried out about 25 years ago. A more recent and reliable upper limit was derived by measuring the depth of the detected soft X-ray shadows cast by high latitude clouds in Ursa Major (Paper 2), but it is about 45% larger than the SMC limit, and most likely contains contribution from the Galactic halo (Snowden et al. 1994b). This uncertainty should be kept in mind in the following discussion.

### 5.1. *Limits on the Unresolved Emission*

The derived intensity limits on 1/4 keV emission can be used to constrain the average emission measure of any X-ray emitting halo in the distant galaxies. The equilibrium thermal emission model of Raymond & Smith (1977; updated in 1993) has been used to derive average emission measure limits from the *ROSAT* PSPC count rate limits in Table 6a and Table 7a, assuming a temperature of $10^6$ K and solar abundances. Fig. 11 summarizes the results. Although it may not be reasonable to expect thermal equilibrium, it is a useful standard for comparison with other work. Other models could easily be used.

The 95% confidence upper and non-negative lower limits to the 1/4 keV emission in the observed galaxies are plotted in dashed lines in Fig. 11. The most probable values for the intensity of the emission we derived with and without correction for shadowing of the *unknown* extragalactic background are shown as solid lines. The gap between the solid lines shows the magnitude of the shadowing correction while the distance between a solid line and the adjacent dashed line indicates the statistical uncertainty. We see significant detection of emission in rings 1 and 2 in all the galaxies, except for ring 1 in NGC 4395, where a bright point source at the center forced the removal of almost all of the central ring. Significant lower limits beyond the second ring were derived only for NGC 5055, and even this should be considered with some reservation. A higher-than-assumed transmission could easily make the lower limits negative -– see equation (1).

NGC 4736 is the brightest of our target galaxies at the center, as can be seen from the lower limits to the emission in Table 7. This galaxy has a very strong central X-ray source, perhaps indicating the presence of some activities of active galactic nuclei (AGN) in NGC 4736 (Cui, Feldkun, & Braun 1996). Although the central source itself has been effectively removed from our analysis, it is possible that the diffuse emission in the central region is influenced by AGN. NGC 4395 is a Seyfert I galaxy (Filippenko, Ho, & Sargent 1993). We detected a bright central source in this galaxy, but it does not show any particularly strong extended central emission (see Table 6 and Table 7).

We replotted the derived upper limits in Fig. 12 along with the local halo surface brightness for three small regions in our own galaxy, derived from two observations toward the Draco Nebula (Snowden et al. 1991; Mendenhall 1993: a re-analysis of Burrows & Mendenhall 1991) and one in Ursa Major (Snowden et al. 1994b). Since these are high-latitude observations, we have plotted them at a position equivalent to the solar radius. The two Draco fields are separated by only a few degrees, but the implied halo intensities in these directions differ by about a factor of 2. Moreover, the halo emission in Ursa Major is at least a factor of 7 lower than the larger Draco result. The patchiness of the halo in our Galaxy is therefore apparent. For comparison, the surface brightness of the Local Bubble as it would appear when observed from outside our own galaxy has been converted to an emission measure and is also plotted on the graph. It lies barely above the Ursa Major point.

In Fig. 12, all emission measure upper limit curves lie below the implied intensity of the Galactic halo emission from the observation of the Draco clouds by Snowden et al. (1991). Therefore, our results severely constrain the covering factor of regions in the observed galaxies with emission measure similar to that of the bright halo in the direction of Draco. On the other hand, the 2 σ upper limit on the emission measure of possible halo emission in the direction of Ursa Major is well below all the curves, so a halo with these average properties in the observed galaxies is always allowed by our data.

In the 0.5-0.9 keV energy range, the deep ROSAT pointing in the Lockman Hole region (H93) resolved ~38% of the diffuse X-ray background observed at high latitudes into primarily extragalactic discrete sources, but the existence of approximately the same observed flux in opaque regions of the Galactic plane shows that there must also be some significant Galactic contribution throughout the disk (McS). This Galactic emission has been modeled as thermal emission from hot plasma at a temperature of ~2-3 $\times 10^6$ K (McS; Garmire et al. 1992). However, its location and true origin are still unknown. Our results on the observed 3/4 keV emission are summarized in Fig. 13.

Fig. 13 was generated the same way as Fig. 11, but assumed an equilibrium thermal spectrum with a temperature of $10^{6.4}$ K. The radial profiles of the 3/4 keV emission are generally very similar to those of the 1/4 keV emission, but appear to be a little more extended in NGC 3184, since it shows definite emission at ring 3 (see Table 10). Also shown on the graph is the upper limit on the surface brightness of the diffuse emission in this energy range from our own galaxy near the solar radius: $150 \times 10^{-6}$ count s$^{-1}$ arcmin$^{-2}$. It is derived by averaging the M-band map of the Wisconsin all-sky survey (McCammon et al. 1983) for galactic latitudes |b| > 60° ($120 \times 10^{-6}$ count s$^{-1}$ arcmin$^{-2}$), then subtracting the contribution of the ROSAT sources (45 in these units), and finally doubling the result to get an upper limit to what one would observe when looking through the entire disk from above. We emphasize that this number probably contains a large extragalactic contribution.

In the 0.9-2.0 keV energy range, the diffuse X-ray background observed in our own galaxy is thought to be mostly extragalactic, but there appears to be strong diffuse emission around the Galactic center. Fig. 14 shows the radial profiles of the observed X-ray intensity in our target galaxies at an effective energy of 1.26 keV, assuming a power law spectrum $\propto E^{-0.4}$, which is a good fit to the observed diffuse background spectrum above 3 keV. The derived differential intensity at the effective energy is not very sensitive to the assumed power law index in the range of 0.4-2.0, but the effective energy does vary somewhat with the assumed spectrum. In this case, the figure shows the measured surface brightness of the emission in this energy band. No assumption about the extragalactic background intensity is required, since the galaxies are effectively transparent at these energies, and no shadowing correction of the unknown extragalactic X-ray background needs to be made. We again averaged the Wisconsin all-sky survey IJ-band maps over galactic latitudes |b| > 60° to derive a surface brightness of $140 \times 10^6$ count s$^{-1}$ arcmin$^{-2}$ near the solar radius, subtracted the contribution of the ROSAT sources (74 in these units), and then doubled it to derive an upper limit on the surface brightness of the diffuse emission from our galaxy when observed from outside. (The actual intensity is probably close to zero at the solar radius.) This result is plotted on the graph as a heavy bar at approximately the equivalent solar radius.

The observed emission radial profiles are similar to those at lower energies, but are more extended: emission at ring 3 is now significant in all of the observed galaxies (see Table 12).

## 5.2. *Filling Factors of $10^6$ K Hot Bubbles*

We can use the intensity limits on 1/4 keV emission to place upper limits on the filling factor of X-ray emitting hot gas in bubbles similar to the one surrounding the Sun that are located within the disks of the distant galaxies. Averaging the Wisconsin sky survey C-band map over galactic latitudes $|b| > 60°$, doubling it, and then absorbing it by an assumed average column density of overlying cool gas of $1.5 \times 10^{20}$ cm$^{-2}$, the Local Bubble surrounding the Sun would have a surface brightness of ~$450 \times 10^{-6}$ count s$^{-1}$ arcmin$^{-2}$ in the R12 band (~405 in the same units for the R12L band) when observed from outside of the Galaxy. Using this number and the derived upper limits on the 1/4 keV emission from Table 6, we derived 95% confidence upper limits on the fractional area of the disks in the observed galaxies that could be occupied by hot bubbles similar to the Local Bubble in our own Galaxy. The results for the inner four rings are shown in Table 13 (numbers in parentheses are the upper limits derived without correction for absorption by gas in the H II layer). The numbers in the table show that the filling factor of hot bubbles is not constrained by our data, except in ring 4 for some of the galaxies.

Table 13: 95% Confidence Upper Limits on the Filling Factors

| Ring | NGC 3184 | NGC 4736 | M101 | NGC 4395 | NGC 5055 |
|---|---|---|---|---|---|
| 1 | 11.56 | 23.86 | 10.54 | 4.41 | 13.59 |
|   | (8.27) | (17.29) | (7.35) | (3.46) | (9.81) |
| 2 | 7.08 | 7.40 | 3.41 | 3.81 | 7.49 |
|   | (5.17) | (5.34) | (2.60) | (3.03) | (5.55) |
| 3 | 3.00 | 2.06 | 1.22 | 1.99 | 3.19 |
|   | (2.33) | (1.66) | (1.12) | (1.73) | (2.54) |
| 4 | 0.35 | 0.94 | 0.69 | 1.06 | 2.38 |
|   | (0.48) | (0.83) | (0.72) | (1.06) | (1.97) |

In M101, McCammon & Sanders (1984) found that the fraction of the disk occupied by hot bubbles similar to the Local Bubble in the Milky Way galaxy was at most 25% in the region between 5′ and 10′ from the center, but the upper limit on the covering factor we now derive in this region (approximately ring 2 + ring 3 in our case) is larger than unity, despite the factor of five improvement in the statistical precision of the current observations. This comes about because McCammon & Sanders did not allow for the possibility of a large extragalactic background, with its attendant correction for shadowing of background flux that might compensate for part or all of the emission. The SMC upper limit we used to correct for the shadowing of the 1/4 keV extragalactic background is about 40% larger than the surface brightness the Local Bubble would have when observed from outside the Galaxy. Since the inner parts of the galaxy are almost opaque, the shadowing correction alone is equivalent to a covering factor of more than unity.

### 5.3. *Limits on Total Energy Radiated by Hot Gas*

We can use our derived upper limits on the X-ray surface brightness of these galaxies in the various energy bands to constrain the total power radiated by hot gas. This is useful, for example, in limiting the avenues for the disposition of supernova energy input. These limits will be strongly dependent on the assumed temperature of the emission, and on the amount of the H I in the observed galaxies that is assumed to overlie the hot gas. We will therefore calculate the limits as a function of these two parameters.

Up to this point, we have discussed the source fluxes outside our own Galaxy in order to facilitate comparisons between observed galaxies with different amount of foreground absorption. This was done by correcting the observed intensities for foreground absorption, using assumed emission temperatures for the observed galaxy that were similar to those observed in our own. Now we wish to place limits over a wide range of temperatures, over which the effective absorption cross section for the 1/4 keV band will vary considerably. We therefore begin by reabsorbing the extragalactic limits (using the originally assumed temperatures) to get limits on observed counting rates. The model fluxes to be compared with them can then readily be corrected for both foreground absorption and the absorption local to the source galaxy due to an assumed overburden of cool gas using the effective cross sections approximate to each temperature and column density.

As before, we have used the equilibrium thermal emission models of Raymond & Smith (1993) with normal abundances to generate X-ray spectra for a range of assumed temperatures. For a given assumed total H I column density (including foreground H I in our own galaxy), we use the *ROSAT* PSPC response function to generate a model curve of the PSPC count rate per unit emission measure in any of the three energy bands versus the temperature. The Raymond & Smith model is also used to generate a curve of total cooling power per unit emission measure versus temperature. The ratio of these is a curve of total cooling power per PSPC count rate in a given energy band versus temperature. We then use these to convert the upper limits on the observed counting rate limits from the galaxies to radiated power limits. This procedure is repeated for the four inner rings in each galaxy.

Fig. 15 show the 95% confidence upper limits on the surface brightness of total radiated power as a function of emission temperature. The limits derived from each band are shown separately for three different values of overlying absorption.

The upper limits in the R12 (or R12L) band provide the strongest constraints at temperatures around $10^6$ K, while those in the R45 and R67 bands are better at higher temperatures. The lower envelope of the three curves for the different energy bands for a given assumed $N_{H\,I}$ therefore gives the best upper limit as a function of temperature. $N_{H\,I} = 0$ is the appropriate curve for limits on halo emission from the near side of the galaxies, while $N_{H\,I} = 3 \times 10^{20}$ H I cm$^{-2}$ represents approximately the half thickness of the disks, and is a reasonable upper limit to the absorption of soft X-rays from hot gas embedded within them.

If we assume a supernova rate of one per 30 years per galaxy, and an average supernova energy of $10^{51}$ ergs, we get an energy dissipation rate of $10^{42}$ ergs s$^{-1}$ through this channel. We then calculate the corresponding average surface brightness of that galaxy, under the assumption that it is uniformly distributed over the galaxy within the optical radius, using the distances to these galaxies in Table 1. The results are shown as dot-dashed lines on the graphs. This allows us to put upper limits to the temperatures of hot gas at each ring in the disks of each galaxy if it were to dissipate an appreciable fraction of the total supernova power (> 30%). These results are given in Table 14.

Table 14: Upper Limits on the Temperature of Hot Gas (°K)

| Ring | NGC 3184 | | NGC 4736 | | M101 | | NGC 4395 | | NGC 5055 | |
|---|---|---|---|---|---|---|---|---|---|---|
| | $N_{H\,I}$ ($10^{20}$ cm$^{-2}$) | | $N_{H\,I}$ ($10^{20}$ cm$^{-2}$) | | $N_{H\,I}$ ($10^{20}$ cm$^{-2}$) | | $N_{H\,I}$ ($10^{20}$ cm$^{-2}$) | | $N_{H\,I}$ ($10^{20}$ cm$^{-2}$) | |
| | 0 | 3 | 0 | 3 | 0 | 3 | 0 | 3 | 0 | 3 |
| 1 | $10^{6.0}$ | $10^{6.2}$ | $10^{6.2}$ | $10^{6.2}$ | $10^{6.3}$ | $10^{6.3}$ | $10^{5.8}$ | $10^{6.2}$ | $10^{6.2}$ | $10^{6.2}$ |
| 2 | $10^{5.9}$ | $10^{6.2}$ | $10^{5.9}$ | $10^{6.1}$ | $10^{5.9}$ | $10^{6.3}$ | $10^{5.8}$ | $10^{6.1}$ | $10^{6.2}$ | $10^{6.2}$ |
| 3 | $10^{5.8}$ | $10^{6.1}$ | $10^{5.7}$ | $10^{6.0}$ | $10^{5.8}$ | $10^{6.0}$ | $10^{5.7}$ | $10^{6.0}$ | $10^{5.9}$ | $10^{6.1}$ |
| 4 | $10^{5.6}$ | $10^{5.8}$ | $10^{5.7}$ | $10^{5.9}$ | $10^{5.7}$ | $10^{6.0}$ | $10^{5.7}$ | $10^{5.9}$ | $10^{5.8}$ | $10^{6.0}$ |

The previous study of M101 by McCammon & Sanders (1984) derived similar upper limits of about $10^{5.7}$ and $10^{5.8}$ K to the temperature of hot gas in this galaxy based on the same assumptions. The upper limit on the halo emission is very close to our results for the rings 3 and 4 in the galaxies for which we have reliable H I data. The upper limits on hot gas embedded in the disks are lower than all our results, again because they did not take into account possible shadowing of the extragalactic background as discussed in the last section.

## 6. CONCLUSIONS

The main results of this paper can be summarized as follows:

1) X-ray emission at 0.1-2.0 keV was detected in all of the observed spiral galaxies, with similar radial profiles for different galaxies and for different energy bands. The emission appears to be a little more radially extended at higher energies.

2) The derived upper limits to the 1/4 keV emission constrain the average emission measure of putative unabsorbed halo emission in the observed galaxies to be no brighter than the apparent halo emission behind the Draco Nebula in our Galaxy (Snowden et al. 1991).

3) The derived upper limits can also be used to constrain the total energy radiated by hot gas as a function of its temperature for various assumed absorbing geometries. If hot gas in the disks of the observed galaxies were to dissipate more than 30% of supernova power, assuming an overlying absorbing column density of $3 \times 10^{20}$ cm$^{-2}$ and a supernova rate of one per 30 years per galaxy, the emission temperature should be less than about $10^{6.3}$ K inside the solar radius, and less than about $10^{6.1}$ K outside. The corresponding upper limits on the emission temperatures of unabsorbed hot halos are much tighter: $\sim 10^{5.9}$ K outside the solar radius (see Table 14).

We would like to thank R. Braun for providing the 21 cm data on NGC 4736 in advance of publication and for useful discussions, J. Kamphuis for providing the M101 H I data in digital format and R. Sancisi for useful discussions. We are grateful to D. P. Cox for many valuable suggestions and discussions on the content and structure of the paper, and to R. Edgar and R. Smith for providing the results of the Raymond-Smith model. We also wish to thank the referee, Daniel Wang, for comments that resulted in an improved manuscript. In determining the long-term enhancement contamination, we partly relied on the ROSAT all-sky survey data provided to us courtesy of Max-Planck-Institut für Extraterrestrische Physik. We have also made use of the NASA/IPAC Extragalactic Database and the HEASARC archival database. This work has been supported in part by NASA grants NAG5-629 and NAG5-1665. Support for D. S. W. was provided by NASA through grant number HF-1040.01-92A from the Space Telescope Science Institute, which is operated by the Association of Universities for Research in Astronomy, Inc., under NASA contract NAS5-26555.

FIGURE CAPTIONS

Fig. 1.—Cleaned and smoothed 1/4 keV X-ray image of NGC 4736. The image is 128 × 128 with a pixel size of 14.947″. The white and black pixels are 0 and $2000 \times 10^{-6}$ count s$^{-1}$ arcmin$^{-2}$, respectively. The small circles show the cutout areas around the point sources detected while the big one in the middle indicates the optical extent of this galaxy.

Fig. 2.—Same as Fig. 1, but for the 3/4 keV band. The white and black pixels are 0 and $400 \times 10^{-6}$ count s$^{-1}$ arcmin$^{-2}$, respectively.

Fig. 3.—Same as Fig. 1, but for the 1.5 keV band. The white and black pixels are 0 and $600 \times 10^{-6}$ count s$^{-1}$ arcmin$^{-2}$, respectively.

Fig. 4.—Radial profiles of unresolved 1/4 keV X-ray emission corrected for the foreground absorption (arbitrary zero). The dashed line shows the off-galaxy baseline level.

Fig. 5.—Radial profiles of average 1/4 keV X-ray transmission from interferometric H I data (see text for references). The assumed transmission curves for NGC 4395 and NGC 5055 are also shown.

Fig. 6.—Corrected radial profiles of unresolved 1/4 keV X-ray emission (arbitrary zero). The dashed line shows the corrected off-galaxy baseline level.

Fig. 7.—Radial profiles of average 3/4 keV X-ray transmission from interferometric H I data (see text for references). The assumed transmission curves for NGC 4395 and NGC 5055 are also shown.

Fig. 8.—Corrected radial profiles of unresolved 3/4 keV X-ray emission (arbitrary zero). The dashed line shows the corrected off-galaxy baseline level.

Fig. 9.—Radial profiles of unresolved 1.5 keV X-ray emission (arbitrary zero). The dashed line shows the off-galaxy baseline level.

Fig. 10.—Same as in Fig. 9, corrected for the effects of non-uniform source removal and the shadowing of the *known* extragalactic background.

Fig. 11.—Emission measure for the X-ray emitting hot gas in the observed galaxies. The most probable values with and without correction for the shadowing of *unknown* extragalactic background are shown in solid lines while the 95% confidence upper and lower limits in dashed lines. The Raymond & Smith (1993) model has been used with a temperature of $10^6$ K and solar abundances to get conversion factors of about 1.585 and 1.515 × 10$^5$ (10$^{-6}$ count s$^{-1}$ arcmin$^{-2}$ per unit emission measure) in the R12 and R12L bands, respectively.

Fig. 12.—Same as in Fig. 11, but only the upper limits are shown. The results for halo emission in our own galaxy derived from observations in Draco and Ursa Major are plotted at the solar radius (the error bar for the Ursa Major result is smaller than the size of the symbol). The open square just above the Ursa Major point indicates the apparent emission measure that a region with the surface brightness of the Local Bubble would have when observed from outside of the Galaxy. The Sun would be approximately at the boundary between the second and third rings.

Fig. 13.—Same as in Fig. 11, but for the 3/4 keV band. An equilibrium thermal spectrum with a temperature of 10$^{6.4}$ K was assumed. The derived conversion factor is about 1.767 × 10$^4$ (10$^{-6}$ count s$^{-1}$ arcmin$^{-2}$ per unit emission measure) in this energy band. The heavy bar at the solar radius indicates an extreme upper limit to the surface brightness our galaxy would have in this energy range when observed from outside (see text).

Fig. 14.—As in Fig. 13, but for the 1.5 keV band. The intensity is given at the effective energy of 1.26 keV, assuming a power law spectrum $\propto E^{-0.4}$, which is a good fit to the observed diffuse background spectrum above 3 keV. The results are not very sensitive to the power law index assumed in the range of 0.4-2.0 if evaluated at the effective energy, but the effective energy varies somewhat with the assumed source spectrum. The derived conversion factor is about 134.3 (10$^{-6}$ count s$^{-1}$ arcmin$^{-2}$ per unit X-ray intensity at 1.26 keV).

Fig. 15a.—95% confidence upper limits to the total energy emitted from hot gas in NGC 3184 as a function of temperature and overlying absorbing column density for the inner four rings. The limits derived from each energy band are shown separately for three different assumptions for the overlying absorbing column. The assumed spectra are those of Raymond & Smith (1993). The dot-dashed line shows the average surface brightness expected if 10$^{42}$ ergs s$^{-1}$ were being radiated by gas within the outer edge of the fourth ring of the galaxy.

Fig. 15b.—Same as Fig. 15a, for NGC 4736.

Fig. 15c.—Same as Fig. 15a, for M101.

Fig. 15d.—Same as Fig. 15a, for NGC 4395.

Fig. 15e.—Same as Fig. 15a, for NGC 5055.